\documentclass[prl,preprintnumbers,aps,twocolumn,nofootinbib]{revtex4-2}
\usepackage{amsmath} 
\usepackage{amssymb}
\usepackage{graphicx,epsfig}
\usepackage[mathscr]{eucal}
\usepackage{color}
\usepackage{natbib}
\usepackage[english]{babel}
\usepackage{tabularx}
\usepackage{setspace}
\usepackage{xspace}
\usepackage{multirow}
\usepackage[usenames,dvipsnames]{xcolor}

\newcommand\as{\alpha_{\mathrm{S}}}

\def\to{\rightarrow} 
\def\nn{\nonumber}

\usepackage{booktabs,cellspace}

\usepackage{scalefnt,pstricks}

\def\rcut{r_{\rm cut}}

\def\ttW{\ensuremath{t \bar t W}\xspace}
\def\ttWm{\ensuremath{t \bar t W^-}\xspace}
\def\ttWp{\ensuremath{t \bar t W^+}\xspace}
\def\ttWpm{\ensuremath{t \bar t W^\pm}\xspace}
\def\ttH{\ensuremath{t \bar t H}\xspace}
\def\bbW{\ensuremath{b \bar b W}\xspace}
\def\Wbb{\bbW}

\def\Hn{\ensuremath{H^{(n)}}\xspace}

\def\ttb{\ensuremath{t {\bar t}}\xspace}
\def\tttt{\ensuremath{t {\bar t} t {\bar t}}\xspace}

\newcommand\Matrix{{\sc Matrix}\xspace}
\newcommand\OpenLoops{{\sc OpenLoops}\xspace}
\newcommand\Recola{{\sc Recola}\xspace}

\newcommand\Whizard{{\sc Whizard}\xspace}

\usepackage{array}
\newcolumntype{L}[1]{>{\raggedright\let\newline\\\arraybackslash\hspace{0pt}}m{#1}}
\newcolumntype{C}[1]{>{\centering\let\newline\\\arraybackslash\hspace{0pt}}m{#1}}
\newcolumntype{R}[1]{>{\raggedleft\let\newline\\\arraybackslash\hspace{0pt}}m{#1}}

\usepackage[colorlinks=true,allcolors={blue!70!black}]{hyperref}

\begin{document} 
\preprint{
ZU-TH 27/23,
TIF-UNIMI-2023-12,
PSI-PR-23-19
}

\title{Precise predictions for the associated production of \\ a \texorpdfstring{$\boldsymbol{W}$}{W} boson with a top-antitop quark pair at the LHC}


\author{Luca Buonocore$^{a}$, Simone Devoto$^{b}$, Massimiliano Grazzini$^{a}$, Stefan Kallweit$^{c}$,\\[0.6ex]
Javier Mazzitelli$^{d}$, Luca Rottoli$^{a}$ and Chiara Savoini$^{a}$\vspace{1em}}

\affiliation{(a) Physik Institut, Universit\"at Z\"urich, CH-8057 Z\"urich, Switzerland,\\
(b) Dipartimento di Fisica, Universit\`a degli Studi di Milano, and INFN, Sezione di Milano, I-20133 Milano, Italy\\
(c) Dipartimento di Fisica, Universit\`{a} degli Studi di Milano-Bicocca and INFN, Sezione di Milano-Bicocca, I-20126 Milano, Italy,\\
(d) Paul Scherrer Institut, CH-5232 Villigen PSI, Switzerland}

\begin{abstract}

\noindent
The production of a top-antitop quark pair in association with a $W$ boson~($\ttW$) is one of
the heaviest signatures currently probed at the Large Hadron Collider~(LHC).
Since the first observation reported in 2015 the corresponding rates have
been found to be consistently higher than the Standard Model predictions,
which are based on next-to-leading order~(NLO) calculations in the QCD and
electroweak~(EW) interactions.
We present the first next-to-next-to-leading order~(NNLO) QCD computation of
\ttW production at hadron colliders. The calculation is exact,
except for the finite part of the two-loop virtual corrections, which is
estimated using two different approaches that lead to consistent results within
their uncertainties. We combine the newly computed NNLO QCD corrections with the
complete NLO QCD+EW results, thus obtaining the most advanced perturbative
prediction available to date for the \ttW inclusive cross section. The tension
with the latest ATLAS and CMS results remains at the \mbox{$1\sigma-2\sigma$} level.

\end{abstract}

\maketitle

\paragraph{Introduction.}
The final state of a $W^\pm$ boson produced in association with a
top-antitop quark pair~($\ttW$) represents one of the most massive Standard Model~(SM)
signatures accessible at the Large Hadron Collider~(LHC).
Since the top quarks rapidly decay into a $W$ boson and a $b$ quark,
the \ttW process leads to two $b$ jets and three decaying $W$ bosons.
This in turn gives rise to multi-lepton signatures that are relevant to a
number of searches for physics beyond the Standard Model~(BSM). In particular,
\ttW production is one of the few SM processes that provides an irreducible
source of same-sign dilepton pairs.
Additionally, the \ttW signature is a relevant background
for the measurement of Higgs boson production in association with a top-antitop quark
pair~(\ttH) and for four-top~(\tttt) production.

Measurements of \ttW production carried out by the ATLAS and CMS collaborations
at centre-of-mass energies of \mbox{$\sqrt{s}=8$\,TeV}~\cite{ATLAS:2015qtq,CMS:2015uvn} and
\mbox{$\sqrt{s}=13$\,TeV}~\cite{ATLAS:2016wgc,CMS:2017ugv,ATLAS:2019fwo} lead to
rates consistently higher than the SM predictions.
A similar situation holds for \ttW measurements in the context of
\ttH~\cite{ATLAS:2019nvo,CMS:2020mpn} and \tttt~\cite{CMS:2019rvj,ATLAS:2020hpj}
analyses.
The most recent measurements~\cite{CMS:2022tkv,ATLAS:2023gon}, based on an
integrated luminosity of about 140 ${\rm fb}^{-1}$, confirm this picture,
with a slight excess at the \mbox{$1\sigma-2\sigma$} level.

In this context, it is clear that the availability of precise theoretical predictions for
the \ttW SM cross section is of the utmost importance.
The next-to-leading order~(NLO) QCD corrections to \ttW production have been
computed in Refs.~\cite{Badger:2010mg,Campbell:2012dh,Maltoni:2015ena}, and EW corrections
in Refs.~\cite{Frixione:2015zaa,Frederix:2017wme}.
Soft-gluon effects were included in Refs.~\cite{Li:2014ula,Broggio:2016zgg,Kulesza:2018tqz,Broggio:2019ewu}.
NLO QCD effects to the complete off-shell \ttW process have been considered in
Refs.~\cite{Bevilacqua:2020pzy,Denner:2020hgg,Bevilacqua:2020srb},
while the complete off-shell NLO QCD+EW computation was reported in Ref.~\cite{Denner:2021hqi}.
Very recently, even NLO QCD corrections
to off-shell \ttW production in association with a light jet were computed~\cite{Bi:2023ucp}.
A detailed investigation of theoretical uncertainties for multi-lepton \ttW signatures
has been presented in Ref.~\cite{Bevilacqua:2021tzp} (see also Ref.~\cite{FebresCordero:2021kcc}).
Current experimental measurements are compared with NLO QCD+EW predictions supplemented
with multijet merging~\cite{Frederix:2012ps,Frederix:2021agh},
which are still affected by relatively large uncertainties.
To improve upon the current situation, next-to-next-to-leading order~(NNLO) QCD
corrections are necessary.

In this Letter we present the first computation of \ttW production at NNLO in QCD.
While the required tree-level and one-loop
scattering amplitudes can be evaluated with automated tools, the two-loop amplitude for
\ttW production is yet unknown. In this work, we estimate it by using two different
approaches. The first parallels the approach successfully applied
in Ref.~\cite{Catani:2022mfv} to \ttH production, and
is based on a {\it soft-$W$ approximation}, which allows us to extract the
\ttW amplitude from the
two-loop amplitudes for top-pair production~\cite{Barnreuther:2013qvf} (see also Ref.~\cite{Mandal:2022vju}). The second
is based on the NNLO calculation of Ref.~\cite{Buonocore:2022pqq}, where an
approximate form of the two-loop amplitude for the production of a heavy-quark pair and
a $W$ boson is obtained from the leading-colour two-loop amplitudes for a $W$ boson and four massless partons~\cite{Badger:2021nhg,Abreu:2021asb} through a {\it massification} 
procedure~\cite{Penin:2005eh,Mitov:2006xs,Becher:2007cu}.
We demonstrate that the two approximations, despite their distinct conceptual foundations and the fact that they are used in a regime where their validity is not granted,
yield consistent results within their respective uncertainties.
Finally, we combine the computed NNLO QCD corrections with the complete NLO QCD+EW result,
thus obtaining the most accurate theoretical prediction for this process available to date.

\paragraph{Calculation.}
The QCD cross section for \ttW production can be written as
$\sigma=\sigma_{\rm LO}+\Delta\sigma_{\rm NLO}+\Delta\sigma_{\rm NNLO}+...$,
where $\sigma_{\rm LO}$ is the LO cross section, $\Delta\sigma_{\rm NLO}$ the NLO QCD
correction, $\Delta\sigma_{\rm NNLO}$ the NNLO QCD contribution, and so forth.

In addition to the inherent challenges involved in obtaining the relevant scattering
amplitudes, the implementation of a complete NNLO calculation is a difficult task because
of the presence of infrared~(IR) divergences at intermediate stages of the calculation.
In this work NNLO IR singularities are handled and cancelled by using the $q_T$ subtraction formalism~\cite{Catani:2007vq}, extended to
heavy-quark production in
Refs.~\cite{Bonciani:2015sha,Catani:2019iny,Catani:2019hip}.
According to the $q_T$ subtraction formalism, the differential cross section $d\sigma$
can be evaluated as
\begin{equation}
  \label{eq:main}
d\sigma={\cal H}\otimes d\sigma_{\rm LO}+\left[d\sigma_{\rm R}-d\sigma_{\rm CT}\right]\, .
\end{equation}
The first term on the right-hand side of Eq.~(\ref{eq:main}) corresponds to the \mbox{$q_T=0$}
contribution. It is obtained through a convolution, with
respect to the longitudinal-momentum fractions $z_1$ and $z_2$ of the colliding partons,
of the perturbatively computable function ${\cal H}$ with the LO cross section $d\sigma_{\rm LO}$.
The real contribution $d\sigma_{\rm R}$ is obtained by evaluating the cross section to
produce the \ttW system accompanied by additional QCD radiation that provides a recoil
with finite transverse momentum $q_T$.
When $d\sigma$ is evaluated at NNLO, $d\sigma_{\rm R}$
is obtained through an NLO calculation by using the dipole subtraction
formalism~\cite{Catani:1996jh,Catani:1996vz,Catani:2002hc}.
The role of the counterterm $d\sigma_{\rm CT}$ is to cancel the singular behaviour of
$d\sigma_{\rm R}$ in the limit \mbox{$q_T\to 0$},
rendering the square bracket term in Eq.~(\ref{eq:main}) finite. The explicit form of
$d\sigma_{\rm CT}$ is completely known up to NNLO: it is obtained by perturbatively
expanding the resummation formula of the logarithmically enhanced contributions to
the $q_T$ distribution
of the \ttW system~\cite{Zhu:2012ts,Li:2013mia,Catani:2014qha,Catani:2021cbl,Ju:2022wia}.

Our computation is implemented within the \Matrix framework~\cite{Grazzini:2017mhc},
suitably extended to \ttW production, along the lines of what was done for heavy-quark
production~\cite{Catani:2019iny,Catani:2019hip,Catani:2020kkl}.
The method was recently applied also to the NNLO calculation of \ttH~\cite{Catani:2022mfv}
and \bbW~\cite{Buonocore:2022pqq} production, for which the contributions from
soft-parton emissions at low transverse momentum~\cite{Catani:2023tby} had to be
properly extended to more general kinematics~\cite{inprep}.
The required tree-level and one-loop amplitudes are obtained with
\OpenLoops~\cite{Cascioli:2011va, Buccioni:2017yxi,Buccioni:2019sur}
and \Recola~\cite{Actis:2016mpe,Denner:2017wsf,Denner:2016kdg}.
In order to  numerically evaluate the contribution in the square bracket of
Eq.~(\ref{eq:main}), a technical cut-off $\rcut$ is introduced on the dimensionless
variable $q_T/Q$, where $Q$ is the invariant mass of the \ttW system.
The final result, which corresponds to the limit \mbox{$\rcut\to 0$}, is extracted by computing
the cross section at fixed values of $\rcut$ and performing the \mbox{$\rcut\to 0$}
extrapolation. More details on the procedure and its uncertainties can be found in
Refs.~\cite{Grazzini:2017mhc,Catani:2021cbl}.

The purely virtual contributions enter the first term on the right-hand side of
Eq.~(\ref{eq:main}), and more precisely the hard function $H$ (related to ${\cal H}$
through \mbox{${{\cal H}=H\delta(1-z_1)\delta(1-z_2)+\delta{\cal H}}$}) whose coefficients,
in an expansion in powers of the QCD coupling $\as(\mu_R)$, are defined as
\begin{equation}\label{eq:Hn}
    \Hn = \frac{2{\rm Re}\left({\cal M}^{(n)}_{\rm fin}(\mu_R, \mu_{\rm IR}) {\cal M}^{(0)*} \right)}{\left|{\cal M}^{(0)}\right|^2}\biggl|_{\mu_R=Q}\,.
\end{equation}
Here, $\mu_R$ is the renormalisation scale, and ${\cal M}_{\rm fin}^{(n)}$
are the perturbative coefficients of the finite part of the renormalised virtual amplitude
for the process \mbox{$u \bar{d}(d \bar{u}) \to t \bar{t} W^{+(-)}$},
after the subtraction of IR singularities at the scale $\mu_{\rm IR}$, according to the
conventions of Ref.~\cite{Ferroglia:2009ii}.
In order to obtain an approximation of the NNLO coefficient $H^{(2)}$, we use two
independent approaches, applied to both the numerator and the denominator of Eq.~(\ref{eq:Hn}).
The first relies on a soft-$W$ approximation.
In the high-energy limit, in which the colliding quark and antiquark of momenta
$p_1$ and $p_2$ radiate a soft $W$ boson with momentum $k$ and polarisation $\varepsilon(k)$,
the multi-loop QCD amplitude in \mbox{$d=4-2\epsilon$} dimensions behaves as
\begin{eqnarray}
  \label{eq:softfact}
  {\cal M}(\{p_i\},k;\mu_R,\epsilon) &\sim &
  \frac{g}{\sqrt{2}} \left( \frac{p_2\cdot \varepsilon^*(k)}{p_2\cdot k}- \frac{p_1\cdot \varepsilon^*(k)}{p_1\cdot k} \right)\nn\\
  &&\times \, {\cal M}_{L}(\{p_i\};\mu_R,\epsilon)  \, ,
\end{eqnarray}
where $g$ is the EW coupling and
${\cal M}_{L}(\{p_i\})$ the \mbox{$q_L{\bar q}_R \to t{\bar t}$} virtual amplitude.
In the second approach the two-loop coefficient $H^{(2)}$ is approximated
in the ultra-relativistic limit \mbox{$m_t \ll Q$}
by using a massification procedure~\cite{Penin:2005eh,Mitov:2006xs,Becher:2007cu}.
We start from the massless \mbox{$W+4$}-parton amplitudes ${\cal M}^{m_t=0}$ evaluated in the
leading-colour approximation~\cite{Abreu:2021asb,Chicherin:2021dyp} to obtain 
\begin{eqnarray}
    \label{eq:massfact}
  {\cal M}(\{p_i\},k;\mu_R,\epsilon)&\sim& Z_{[q]}^{(m_t|0)}(\as(\mu_R),\tfrac{m_t}{\mu_R},\epsilon)  \nn 
  \\ && \times \,
  {\cal M}^{m_t=0}(\{p_i\},k;\mu_R,\epsilon) \, ,
\end{eqnarray}
where $Z$ are perturbative functions whose explicit expression up to NNLO can be found in Ref.~\cite{Mitov:2006xs}.
This procedure\footnote{We note that at two-loop order the massification procedure is
unable to correctly recover the contribution from massive top-quark loops. Therefore,
analogous diagrams in the real-virtual contributions are omitted for consistency.
Accordingly, we do not include the real subprocess with four top quarks entering at NNLO.
We have verified that the latter, which constitutes an estimate of the
impact of the neglected diagrams, has a negligible effect on our results.} was
successfully applied to evaluate NNLO corrections to \Wbb production in
Ref.~\cite{Buonocore:2022pqq}.

In order to use Eq.~(\ref{eq:softfact}) to approximate the \ttW amplitudes, we need
to introduce a prescription that, from an event containing a \ttb pair and a
$W$ boson, defines a corresponding event in which the $W$ boson is removed. This is
accomplished by absorbing the $W$ momentum into the top quarks, thus
preserving the invariant mass of the event. 
On the other hand, for the application of Eq.~(\ref{eq:massfact}) we 
map the momenta of the massive top quarks into massless momenta by
preserving the four-momentum of the \ttb pair.
In both cases we reweight the respective two-loop coefficients with the exact Born matrix elements. This approach effectively captures additional kinematic effects, which we expect to extend the region of validity of the approximations well beyond where it may be assumed in the first place.

For our numerical studies, we consider the on-shell production of a $W$ boson in
association with a \ttb pair in proton collisions, at a centre-of-mass energy
of \mbox{$\sqrt{s}=13$\,TeV}.
We set the pole mass of the top quark to \mbox{$m_t = 173.2$\,GeV}, while for the $W$ mass
we use \mbox{$m_W = 80.385$\,GeV}. We work in the $G_\mu$-scheme for the EW parameters,
with \mbox{$G_\mu = 1.16639\times 10^{-5}$\,GeV$^{-2}$} and \mbox{$m_Z = 91.1876$\,GeV}.
We consider a diagonal CKM matrix.
We use the \verb|NNPDF31_nnlo_as_0118_luxqed| set for parton
distribution functions~(PDF)~\cite{Bertone:2017bme} and strong coupling, which is based on
the \verb|LUXqed| methodology~\cite{Manohar:2016nzj}
to determine the photon density. We adopt the \verb|LHAPDF|
interface~\cite{Buckley:2014ana} and use {\sc PineAPPL}~\cite{Carrazza:2020gss} grids 
through the new {\sc Matrix}$+${\sc PineAPPL} interface~\cite{PineAPPLMATRIX} to estimate PDF and $\as$
uncertainties. 
For our central predictions we set the 
renormalization~($\mu_R$) and factorization~($\mu_F$) scales to the value
\mbox{$\mu_0 = m_t + m_W/2 \equiv M/2$}, and evaluate the scale uncertainties by performing
a 7-point variation, varying them independently by a factor of two with the constraint
\mbox{$1/2 \leq \mu_R/\mu_F \leq 2$}. 

\begin{figure}[t]
  \includegraphics[width=0.49\textwidth]{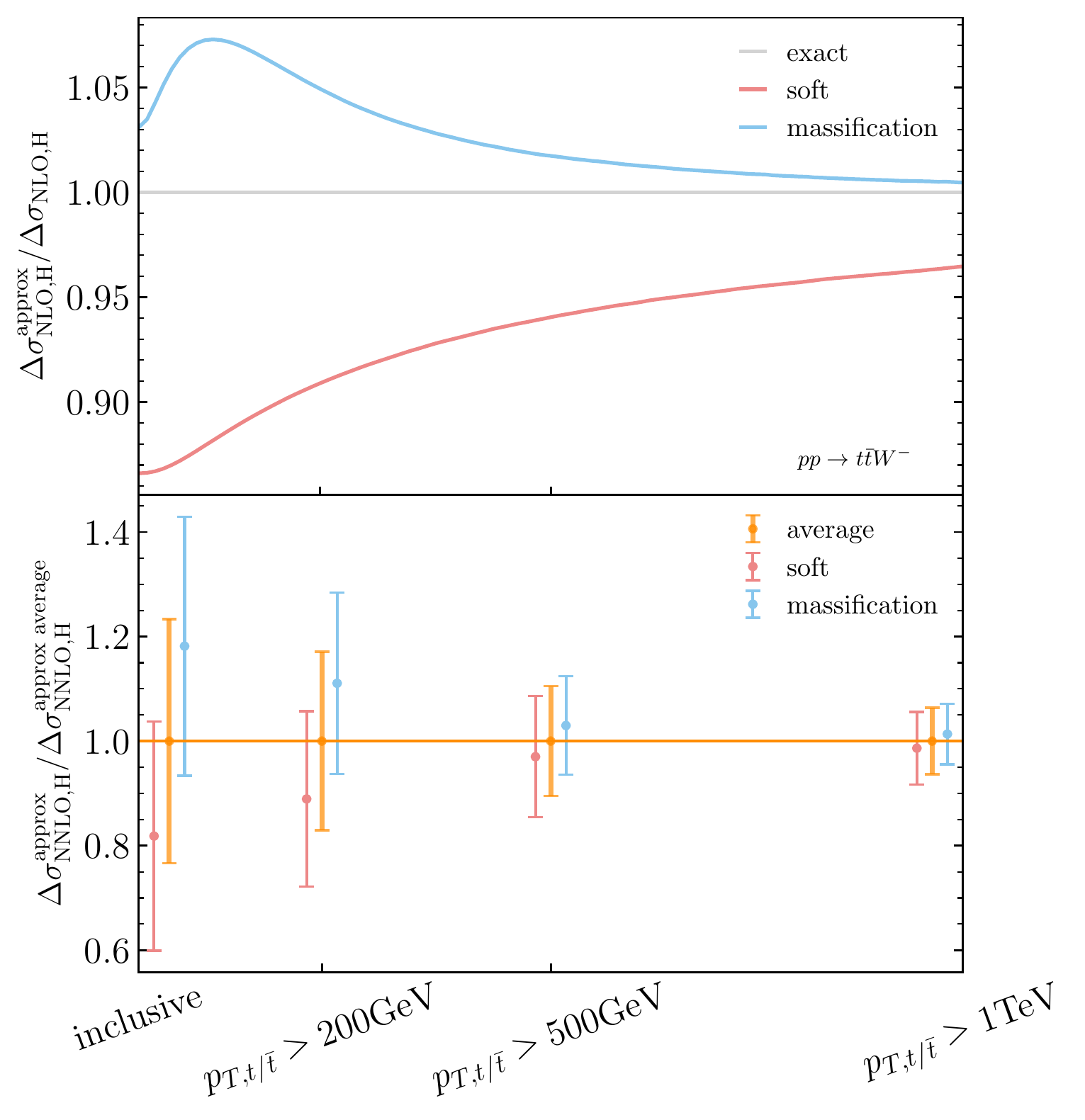}
  \caption{Results for $\Delta\sigma_{\rm NLO,H}$ (upper panel) and $\Delta\sigma_{\rm NNLO,H}$
    (lower panel), in the case of \ttWm production, obtained with the two approximations presented in this work, for different cuts on
    the transverse momenta of the top quarks. At NLO the approximations
    are normalized to the exact result, while at NNLO to their average. The uncertainties of
    each approximation at NNLO are presented, as well as their combination. Similar results
    are obtained for \ttWp.}
    \label{fig:H1_and_H2}
\end{figure}

In order to test the quality of our approximations, we apply them to evaluate the
contribution of the coefficient $H^{(1)}$ to the NLO correction, $\Delta\sigma_{\rm NLO,H}$.
In Fig.~\ref{fig:H1_and_H2} (upper panel) the two approximations are compared to the exact
result, as functions of the cut on the transverse momenta of the top quarks,
$p_{T,t/\bar{t}}$.
We observe that both approximations get closer to the exact result if a harder cut is
imposed, since the large-$p_{T,t/\bar{t}}$ region corresponds to a kinematical configuration
where both of them are expected to reproduce the full amplitude.
In particular, we observe that the soft approximation tends to undershoot the exact result,
while the massification approach overshoots it. Remarkably, both approaches provide a good
approximation also at the inclusive level.

We now move on to the contribution of the coefficient $H^{(2)}$ to the NNLO correction,
$\Delta\sigma_{\rm NNLO,H}$. In Fig.~\ref{fig:H1_and_H2} (lower panel) the two
approximations are compared, normalised to their average. The uncertainties of the soft
and massification results are also depicted. These are evaluated starting from the assumption that
the uncertainty of each approximation of $\Delta\sigma_{\rm NNLO,H}$
is not smaller
than the relative difference between $\Delta\sigma_{\rm NLO,H}^{\rm approx}$ and the exact
NLO result. 
We obtain a first estimate of the uncertainty on $\Delta\sigma_{\rm NNLO,H}$ by conservatively multiplying $\Delta\sigma_{\rm NLO,H}^{\rm approx}$ by a factor of two.
As an additional estimate, we consider variations of the subtraction scale $\mu_{\rm IR}$, at which our approximations are 
applied, by a factor of two around the central scale $Q$ (adding the exact evolution from $\mu_{\rm IR}$ to $Q$).
For each of the two approximations, the uncertainty is defined as the maximum between
these two estimates.
From Fig.~\ref{fig:H1_and_H2} we see that the two approximations are consistent within their respective uncertainties.
We therefore conclude that our approach can provide a good estimate of the true NNLO hard-virtual contribution.
Our best prediction for  $\Delta\sigma_{\rm NNLO,H}$ is finally obtained by taking the average
of the two approximations and linearly combining their uncertainties.
We note that with such procedure the central values of the two approximations are enclosed within the uncertainty band of the average result.
The final uncertainty on $\Delta\sigma_{\rm NNLO,H}$ turns out to be at the
${\cal O}(25\%)$ level.\footnote{We note that a similar control on the two-loop contribution
is obtained in recent calculations for massless \mbox{$2\to 3$} processes employing the leading-colour
approximation (see e.g. Ref.~\cite{Abreu:2023bdp}).}
As we will observe in what follows, this leads to an uncertainty of the NNLO prediction which
is significantly smaller than the residual perturbative uncertainties.

\begin{table*}[t]
\centering
\renewcommand{\arraystretch}{1.5}
\setlength{\tabcolsep}{0.5em}
\begin{tabular}{cllll}
    &
  \multicolumn{1}{c}{$\sigma_{\ttWp}\, [{\rm fb}]$} &
  \multicolumn{1}{c}{$\sigma_{\ttWm}\, [{\rm fb}]$} &
  \multicolumn{1}{c}{$\sigma_{\ttW}\, [{\rm fb}]$} &
  \multicolumn{1}{c}{$\sigma_{\ttWp}/\sigma_{\ttWm}$}
  \\
\hline
  LO$_{\rm QCD}$ &
  $ \phantom{000}283.4^{+25.3\%}_{-18.8\%}$ &
  $ \phantom{000}136.8^{+25.2\%}_{-18.8\%}$ &
  $ \phantom{000}420.2^{+25.3\%}_{-18.8\%}$ &
  $ \phantom{0}2.071^{+3.2\%}_{-3.2\%}$
  \\
  NLO$_{\rm QCD}$ &
  $ \phantom{000}416.9^{+12.5\%}_{-11.4\%}$ &
  $ \phantom{000}205.1^{+13.2\%}_{-11.7\%}$ &
  $ \phantom{000}622.0^{+12.7\%}_{-11.5\%}$ &
  $ \phantom{0}2.033^{+3.0\%}_{-3.4\%}$
  \\
  NNLO$_{\rm QCD}$ &
  $ \phantom{000}475.2^{+4.8\%}_{-6.4\%}\pm 1.9\%$ &
  $ \phantom{000}235.5^{+5.1\%}_{-6.6\%}\pm 1.9\% $ &
  $ \phantom{000}710.7^{+4.9\%}_{-6.5\%} \pm 1.9\%$ &
  $ \phantom{0}2.018^{+1.6\%}_{-1.2\%}$
  \\
  \hline
  NNLO$_{\rm QCD}$+NLO$_{\rm EW}$ &
  $ \phantom{000}497.5^{+6.6\%}_{-6.6\%}\pm 1.8\%$ &
  $ \phantom{000}247.9^{+7.0\%}_{-7.0\%}\pm 1.8\%$ &
  $ \phantom{000}745.3^{+6.7\%}_{-6.7\%}\pm 1.8\%$ &
  $ \phantom{0}2.007^{+2.1\%}_{-2.1\%}$
  \\
  \hline
  ATLAS~\cite{ATLAS:2023gon} &   
  $ \phantom{000}585^{+6.0\%}_{-5.8\%} \phantom{}^{+8.0\%}_{-7.5\%}$ &
  $ \phantom{000}301^{+9.3\%}_{-9.0\%} \phantom{}^{+11.6\%}_{-10.3\%}$ &
  $ \phantom{000}890^{+5.6\%}_{-5.6\%} \phantom{}^{+7.9\%}_{-7.9\%}$ &
  $ \phantom{0}1.95^{+10.8\%}_{-9.2\%} \phantom{}^{+8.2\%}_{-6.7\%}$
  \\
  CMS~\cite{CMS:2022tkv} &   
  $ \phantom{000}553^{+5.4\%}_{-5.4\%} \phantom{}^{+5.4\%}_{-5.4\%}$ &
  $ \phantom{000}343^{+7.6\%}_{-7.6\%} \phantom{}^{+7.3\%}_{-7.3\%}$ &
  $ \phantom{000}868^{+4.6\%}_{-4.6\%} \phantom{}^{+5.9\%}_{-5.9\%}$ &
  $ \phantom{0}1.61^{+9.3\%}_{-9.3\%} \phantom{}^{+4.3\%}_{-3.1\%}$
  \\
\end{tabular}
\caption{\label{tab:xs}Inclusive cross sections for \ttWp and \ttWm production at
different perturbative orders, together with their sum and ratio.
The uncertainties are computed through scale variations and for our best prediction, NNLO$_{\rm QCD}$+NLO$_{\rm EW}$, are symmetrised as discussed in the text. Where NNLO QCD corrections are included, the error from the approximation of the two-loop amplitudes is also shown.
The numerical uncertainties on our predictions are at the per mille level or below. 
The corresponding experimental results from the ATLAS~\cite{ATLAS:2023gon}
and CMS~\cite{CMS:2022tkv} collaborations are also quoted, with their statistical and
systematic uncertainties.}
\end{table*}

\paragraph{Results.}
We now focus on our numerical predictions for the LHC.
Our results for the total \ttWp and \ttWm cross sections are presented in Table~\ref{tab:xs}.
In the first three rows we consider pure QCD predictions, which are labelled
N$^n$LO$_{\rm QCD}$ with \mbox{$n=0,1,2$}. The results in the fourth row, dubbed \mbox{NNLO$_{\rm QCD}$+NLO$_{\rm EW}$},
represent our best prediction. They include additively also EW corrections and all subleading (in $\as$) terms up to NLO,
originally computed in Ref.~\cite{Dror:2015nkp,Frederix:2017wme}. We recompute them here within the \Matrix framework,
after validation against a recent implementation in \Whizard~\cite{Bredt:2022nkq}.
Predictions for the sum and the ratio of the \ttWp and \ttWm cross sections are also
provided, and their scale uncertainties are evaluated by performing 7-point scale variations for each of them,
keeping $\mu_R$ correlated,
while the values of $\mu_F$ for the \ttWp and \ttWm cross sections are allowed to differ by 
at most a factor of two.\footnote{The uncertainty due to the approximation of the two-loop corrections is expected to largely cancel in the ratio.}
Finally, the most recent
results by the ATLAS~\cite{ATLAS:2023gon} and
CMS~\cite{CMS:2022tkv} collaborations are quoted.

We start by discussing the pattern of QCD corrections.
The NLO cross section for both \ttWp and \ttWm production is about $50\%$ larger than the corresponding LO result.
The NNLO corrections are moderate, and increase the NLO result by about $15\%$, showing first signs of perturbative convergence.
The ratio between the two cross sections shows a very stable perturbative behaviour.
The size of the scale uncertainties is substantially reduced at NNLO,
in line with the observed smaller corrections to the central prediction.
The impact of the two-loop contribution is relatively large, about $6\%-7\%$ of the NNLO cross section. 
Nonetheless, we find that the ensuing uncertainty on our prediction is ${\cal O}(\pm 2\%)$,
i.e.\ significantly smaller than the remaining perturbative uncertainties.

\begin{figure}[t]
  \includegraphics[width=0.46\textwidth]{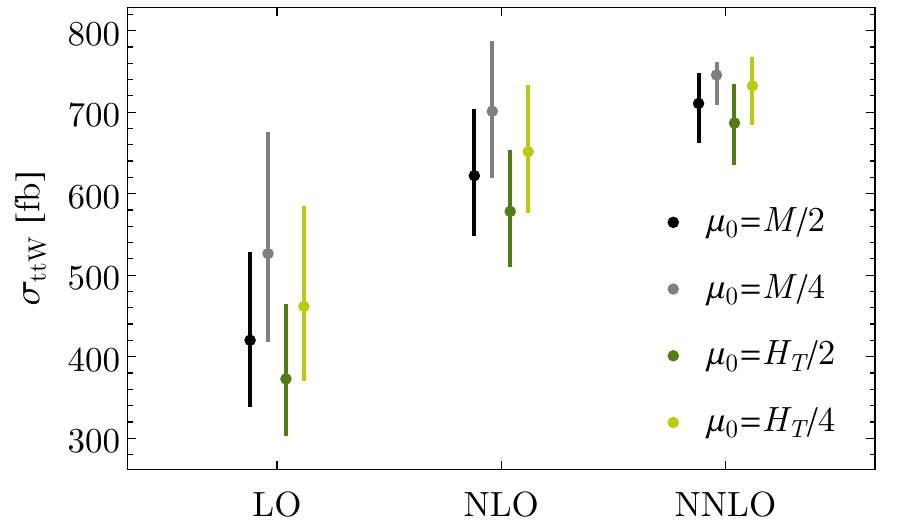}\hspace*{0.02\textwidth}
    \caption{Inclusive \ttW cross sections at different orders in the QCD expansion, for 
    different choices of the central renormalization and factorization scales.}
    \label{fig:qcd_corrections}
\end{figure}

In addition to the value \mbox{$\mu_0=M/2$} used in Table~\ref{tab:xs},
we have also considered alternative choices for the central scale, specifically
\mbox{$\mu_0=M/4$}, $H_T/2$ and $H_T/4$, where 
 $H_T$ is the sum of the transverse masses of the top quarks and the $W$ boson.
Results for the different perturbative orders in the QCD expansion are presented in
Fig.~\ref{fig:qcd_corrections}.
At each order, the four predictions are fully consistent within their uncertainties,
and in particular the \mbox{$\mu_0=M/2$} and \mbox{$\mu_0=H_T/4$} bands cover the central values
of the other scale choices that have been considered.
We note that symmetrising the band of the \mbox{$\mu_0=M/2$} prediction at NNLO leads
to an upper bound which is almost identical to that of the \mbox{$\mu_0=M/4$} and \mbox{$\mu_0=H_T/4$}
scale variations.
Therefore, to be conservative, the perturbative uncertainties affecting our final \mbox{NNLO$_{\rm QCD}$+NLO$_{\rm EW}$} results are estimated by symmetrising the scale variation error.
More precisely, we take the maximum among the upward and downward variations, assign it symmetrically and leave the nominal prediction unchanged.

The EW corrections increase our NNLO$_{\rm QCD}$ cross sections by about $5\%$. 
While smaller than the NNLO QCD corrections, their inclusion is crucial for an 
accurate description of this process, as their magnitude is comparable to the 
NNLO$_{\rm QCD}$ scale uncertainties. 
The PDF ($\as$) uncertainties, not shown in Table~\ref{tab:xs}, on the \ttWp and \ttWm cross 
sections amount to $\pm 1.8\%$ ($\pm 1.8\%$) and $\pm 1.7\%$ ($\pm 1.9\%$), respectively.%
\footnote{We consider 68\% confidence level PDF uncertainties, while the $\as$ uncertainties are computed as half the difference between the corresponding sets for $\as(m_Z)=0.118\pm 0.001$.} The PDF uncertainty on their ratio, derived by recalculating the ratio for each replica, is $\pm 1.7\%$. Its $\as$ uncertainty is negligible.

The current theory reference to which experimental data are compared
is the FxFx prediction of Ref.~\cite{Frederix:2021agh},
which reads \mbox{$\sigma^{\rm FxFx}_{\ttW}=722.4^{+9.7\%}_{-10.8\%}$\,fb}.
Our \mbox{NNLO$_{\rm QCD}$+NLO$_{\rm EW}$} prediction for the \ttW cross section in
Table~\ref{tab:xs} is fully consistent with this value, with considerably smaller uncertainties.

\begin{figure}[t]
  \includegraphics[width=0.46\textwidth]{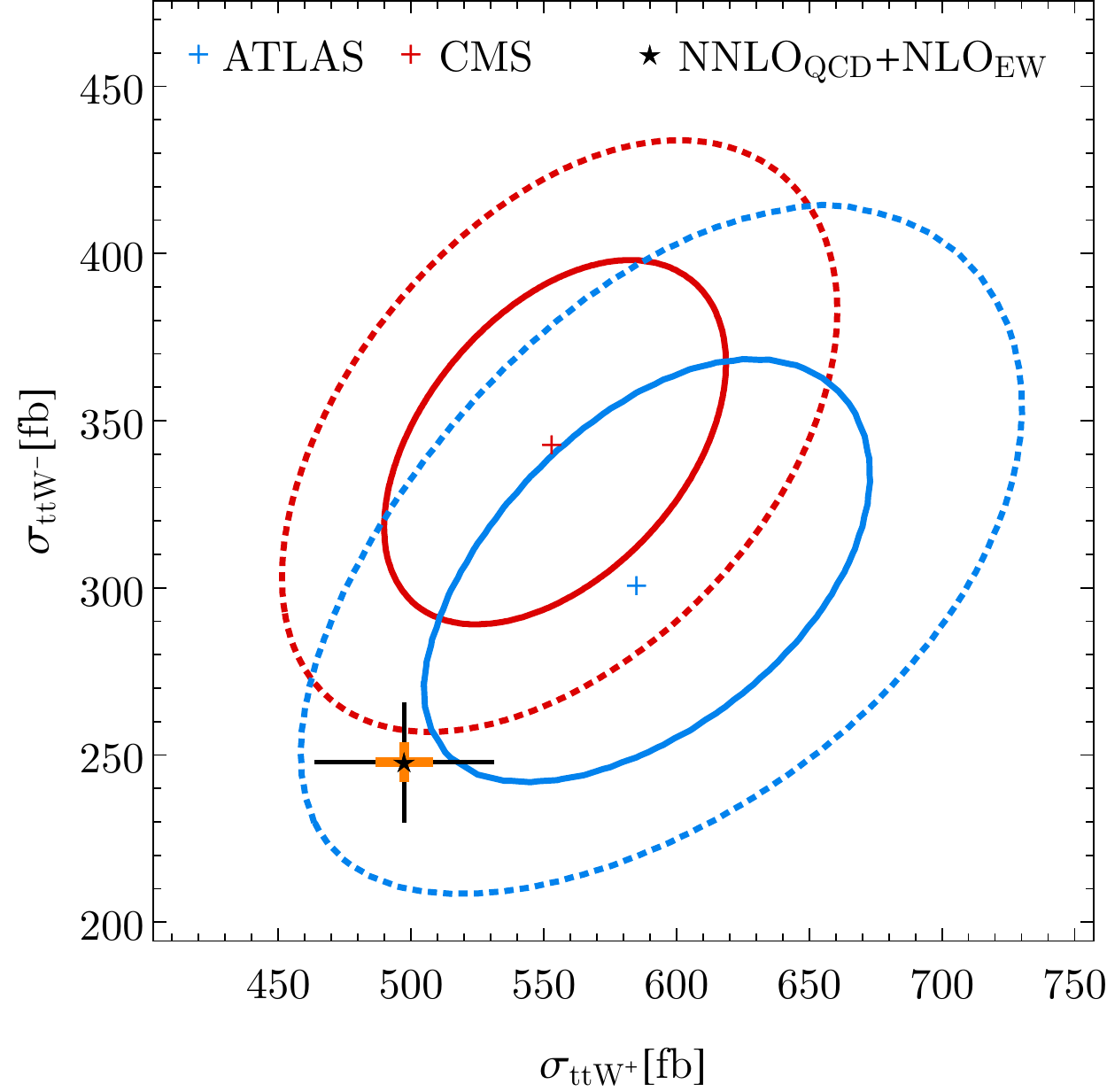}\hspace*{0.02\textwidth}
  \caption{Comparison of our \mbox{NNLO$_{\rm QCD}$+NLO$_{\rm EW}$} result to the measurement
    performed by the CMS (red) and ATLAS (blue) collaborations in
    Refs.\,\cite{CMS:2022tkv,ATLAS:2023gon}, at 68\% (solid) and 95\% (dashed) confidence
    level. We indicate in black and orange the scale and the approximation uncertainties,
    respectively, of the \mbox{NNLO$_{\rm QCD}$+NLO$_{\rm EW}$} result.}
    \label{fig:contour}
\end{figure}

We now compare our theoretical predictions to the measurements performed by the ATLAS and
CMS collaborations in Refs.~\cite{CMS:2022tkv,ATLAS:2023gon}, which represent the most precise
experimental determination of the \ttWpm cross sections to date. 
From Table~\ref{tab:xs} we observe that the individual measurements for the \ttWp and \ttWm
cross sections are systematically above the theoretical predictions, but all within two
standard deviations of our central results, except for the \ttWm measurement by the CMS collaboration.
The measurement of the ratio $\sigma_{\ttWp}/\sigma_{\ttWm}$ by the ATLAS collaboration is
in excellent agreement with our prediction, whereas the CMS result exhibits some tension.

Finally, we present in Fig.~\ref{fig:contour} our \mbox{NNLO$_{\rm QCD}$+NLO$_{\rm EW}$} results with
their perturbative uncertainties in the \mbox{$\sigma_{\ttWp} - \sigma_{\ttWm}$} plane,
together with the 68\% and 95\% confidence level regions obtained by the two collaborations. 
The subdominant uncertainties due to the approximation of the two-loop corrections are also shown.
When comparing to the data, we observe an overlap between the \mbox{NNLO$_{\rm QCD}$+NLO$_{\rm EW}$}
uncertainty bands and the $1\sigma$ and $2\sigma$ contours of the ATLAS and CMS measurements,
respectively.

\paragraph{Summary.}
In this Letter we have presented the first calculation of the second-order QCD corrections to
the hadroproduction of a $W$ boson in association with a top-antitop quark pair.
Our results are exact, except for the finite part of the two-loop virtual corrections, which
is computed by using two independent approximations. While these approximations are completely
different in their conception, they lead to consistent results, thereby providing a strong check of our approach.

We have combined our results with the NLO EW corrections, obtaining the most
precise theoretical determination of the inclusive \ttWpm cross section available to date.
Our results significantly reduce the size of the perturbative uncertainties, allowing for a
more meaningful comparison to the results obtained by the ATLAS and CMS collaborations.
The high level of precision attained by our theoretical predictions will enable even more
rigorous tests of the SM, as more precise experimental measurements become available.

\begin{acknowledgments}

\paragraph{Acknowledgements.}
This work is supported in part by the Swiss National Science Foundation~(SNSF) under
contracts 200020\_188464 and PZ00P2\_201878, by the UZH Forschungskredit Grant FK-22-099, and by the ERC Starting Grant 714788 REINVENT.
We would like to thank Josh McFayden and the ATLAS collaboration for providing us with the data
to produce the contours in Fig.~3.
We are grateful to Pia Bredt for the validation of our EW predictions against her new implementation in \Whizard, and to Jonas Lindert for his ongoing support on {\sc OpenLoops}, in particular for providing several dedicated amplitudes.
We are indebted to Christopher Schwan for his immediate support on {\sc PineAPPL}, including explicit extensions of its functionality.
We thank Stefano Catani, Thomas Gehrmann and Paolo Torrielli for useful discussions,
and Fabio Maltoni for the continuous encouragement to timely pursue this project.

\end{acknowledgments}

\bibliographystyle{UTPstyle} 
\bibliography{biblio}

\end{document}